\documentclass[runningheads,a4paper]{llncs}

\usepackage{amssymb}
\usepackage{graphicx}

\usepackage{url} 
\newcommand{\keywords}[1]{\par\addvspace\baselineskip
\noindent\keywordname\enspace\ignorespaces#1}

\usepackage{hyperref}
\usepackage{float}
\usepackage{caption}
\usepackage{subcaption}
\usepackage{color}

\newcommand{\Eurora}{Eurora} 

\newcommand{\Cineca}{CINECA} 

\newcommand{\revised}[1]{{\color{black}#1}}
\newcommand{\codeurl} 
{\href{http://github.com/alinasirbu/eurora_job_power_prediction}{http://github.com/alinasirbu/eurora\_job\_power\_prediction}}

\newcommand{\codeauth} 
{A. S\^irbu and O. Babaoglu}

\begin{document}

\mainmatter  

\title{Power Consumption Modeling and Prediction in a Hybrid CPU-GPU-MIC Supercomputer}

\titlerunning{Power Consumption Modeling and Prediction}

%
%
\author{Alina S\^irbu$^{1,2}$ \and Ozalp Babaoglu$^2$}
\authorrunning{A. S\^irbu\and O. Babaoglu}

\institute{$^1$ Department of Computer Science, University of Pisa, Italy\\
$^2$ Department of Computer Science and Engineering, University of Bologna, Italy}

\maketitle

\toctitle{Power Consumption Modeling and Prediction in a Hybrid CPU-GPU-MIC Supercomputer}
\tocauthor{Alina S\^irbu \and Ozalp Babaoglu}

\begin{abstract}
Power consumption is a major obstacle for High Performance Computing (HPC) systems in their quest towards the holy grail of ExaFLOP performance. Significant advances in power efficiency have to be made before this goal can be attained and accurate modeling is an essential step towards power efficiency by optimizing system operating parameters to match dynamic energy needs.  In this paper we present a study of power consumption by jobs in {\Eurora}, a hybrid CPU-GPU-MIC system installed at the largest Italian data center. Using data from a dedicated monitoring framework, we build a data-driven model of power consumption for each user in the system and use it to predict the power requirements of future jobs. We are able to achieve good prediction results for over 80\% of the users in the system. For the remaining users, we identify possible reasons why prediction performance is not as good. Possible applications for our predictive modeling results include scheduling optimization, power-aware billing and system-scale power modeling. All the scripts used for the study have been made available on GitHub.
\end{abstract}

\section{Introduction} 

A major impediment for supercomputers from reaching the ExaFLOP target is power consumption. Energy efficiency of computing systems has to increase by at least one order of magnitude to achieve this goal~\cite{bartolini2014}. This requires power optimization at all levels of hardware and software, including computation, networking and cooling. Numerous power modeling studies have been conducted in recent years towards these goals. Models can enable prediction of power usage under different scenarios, and indicate operating modes that optimize energy needs. Optimization can be obtained not only at low levels, e.g. through frequency and voltage scaling present in most modern CPUs, but also at higher levels, e.g. through power-aware scheduling, which has not been extensively studied. 

In this paper, we model power needs of jobs in a hybrid CPU-GPU-MIC system ({\Eurora}) with the aim of predicting power consumption of future jobs before they are started. Among other benefits, accurate prediction could enable development of advanced power-aware schedulers that can optimize power for the same workload. {\Eurora} is a prototype supercomputer that topped the Green500 list in July 2013 for energy efficiency.
It includes an advanced monitoring framework that collects status data in an open-access database~\cite{bartolini2014}, which we use to extract important features that enable prediction of job power consumption. The prediction problem is formulated as a regression task: given feature values (independent variables), compute the power consumption for a job (dependent variable). We divide this task into subproblems corresponding to each component type (CPU, GPU, MIC), and use Support Vector Regression (SVR)~\cite{svr} for each individual problem. The total job power is then obtained as the sum of the individual component power consumptions.

This paper makes several contributions to power research for HPC systems. First, it identifies several features relevant to job power consumption in hybrid systems, supported by data examples.
These include features not previously considered when modeling power
such as application names, same-node resources used by other jobs and job running times.
We model power exclusively through job-related features and do not require knowledge of CPU frequencies, load or application code structure to extract the sequence of executed operations. Second, we build power consumption models \emph{for each user} starting from historical data and employing SVR. Models are shown to have high predictive power for most users.
Third, we perform an analysis of power consumption variability for the system to provide context for the model error levels, and explain model limitations. Finally, we outline a methodology to implement the prediction framework in real time and discuss application scenarios. We used the Google BigQuery data analytics service~\cite{bigquery} for the initial data analysis phase, while model training was done using the \emph{scikit-learn} python package \cite{scikit-learn}. We have made all of the scripts used for our study available on GitHub \cite{scripts-eurora}.


\section{The {\Eurora} system and its data}

{\Eurora} \cite{cavazzoni2012} is a prototype HPC system hosted at {\Cineca} (www.cineca.it) that combines the use of CPUs, GPUs and MICs to achieve higher power efficiency. It remained in production for over 2 years from 2013 to 2015. 
The system consists of 64 nodes, each hosting two 8-core Intel Xeon E5 CPUs and two expansion cards that can contain either GPU or MIC accelerator modules.
There are 3 different classes of CPUs based on their maximum frequencies:
2.1GHz (the \emph{slow} class, denoted as $S$ and present at 24 nodes), 2.2GHz (the \emph{medium} class, denoted as $M$ and present at 8 nodes) and 3.1GHz (the \emph{fast} class, denoted as $F$ and present at 32 nodes). 
Half of the nodes mount GPUs (Nvidia Tesla Kepler) while the other half mount Intel ``Knights Corner'' MIC (Xeon Phi). All nodes run CentOS Linux. The workload is handled through the Portable Batch System (PBS).

{\Eurora} contains an extensive monitoring subsystem  
which collects high resolution (5-second intervals) status data from system components, including power and cooling infrastructures~\cite{bartolini2014}.
Log data for the period 31 March 2014 to 11 August 2015 is available (250GB of data in 328 tables), with several gaps due to system/monitoring errors or database migration operations. The work reported in this paper is based on these data for computing power consumption per job and building a prediction framework for estimating future job power. We limited our study to the data from 2014 since the system underwent several changes in 2015 and became more unstable. 
\emph{Workload information} provided the number of resources used by each job on each node at 5-minute resolution. Although this number is known, it is impossible to extract from the data exactly which CPU/GPU/MIC is being used out of the two available on each node.
\emph{Power logs} allowed us to compute power consumption for each component (CPUs, GPUs and MICs) on each node, again every 5 minutes. 
Power data is known only at the level of CPU, GPU and MICs but is not available for the cores.

Combining workload and power data, we computed at 5-minute intervals the overall power usage of a job $j$ as the sum of the power of each component type: 
\begin{equation}\label{eq_pow}
P^{j} = P^{j}_{S} + P^{j}_{M} + P^{j}_{F} + P^{j}_{GPU} + P^{j}_{MIC}
\end{equation}
Power for each component type is computed by summing over all used nodes. For example, for the $F$ CPU type, 
$
P^{j}_{F}=\sum_{i\in \mathrm{nodes}}P_{F}^{j}(i)
$
where $P_{{F}}^{j}(i)$ is the power used by job $j$ at a fast CPU on node $i$. If the job does not use node $i$ or if CPU type $F$ is not present at node $i$, the corresponding power is assumed to be 0.  Denoting the number of cores used by job $j$ on node $i$ as $n_{j}(i)$ and that used by other jobs as $n_{\mathrm{other}}(i)$, the number of free cores at node $i$  is given by
$n_{\mathrm{idle}}(i)=16-n_{j}(i)-n_{\mathrm{other}}(i)$ (recall that each node contains two 8-core CPUs).  Let $P_{{F}}(i)$ denote the total power recorded for the fast CPU type at node $i$.  Then, the power used by job $j$ at a fast CPU on node $i$ is computed based on the number of cores used by job $j$ in relation to the total number of cores at the node, the number of cores used by other jobs and the total power recorded at the node for that CPU type:
\begin{equation}\label{eq_comp_pow}
P_{{F}}^{j}(i) = n_{j}(i) \frac{P_{F}(i) - \hat{P}_{F} \times n_{\mathrm{idle}}(i)} {n_{j}(i)+n_{\mathrm{other}}(i)}
\end{equation}
where $\hat{P}_{F}$ denotes the average power consumed by a single $F$ type CPU core when it is idle.
We estimate this quantity from the log data by
dividing the total power consumed by an idle CPU of type $F$ by the number of cores (which is 8).
The same procedure is repeated for the remaining types $M$, $S$, GPUs and MICs.
This procedure may introduce some noise in calculating job power, since it assumes that when two jobs share the same node, the power usage is evenly distributed across used components (e.g., cores). It is highly unlikely that this assumption holds since jobs have different power needs, yet it is necessary in order to be able to use the entire job set in our study,
\revised{since many jobs indeed share nodes with other jobs.}
%


\begin{figure}[!b]
  \begin{center}
    \begin{subfigure}{0.48\textwidth}
    \includegraphics[width=\textwidth]{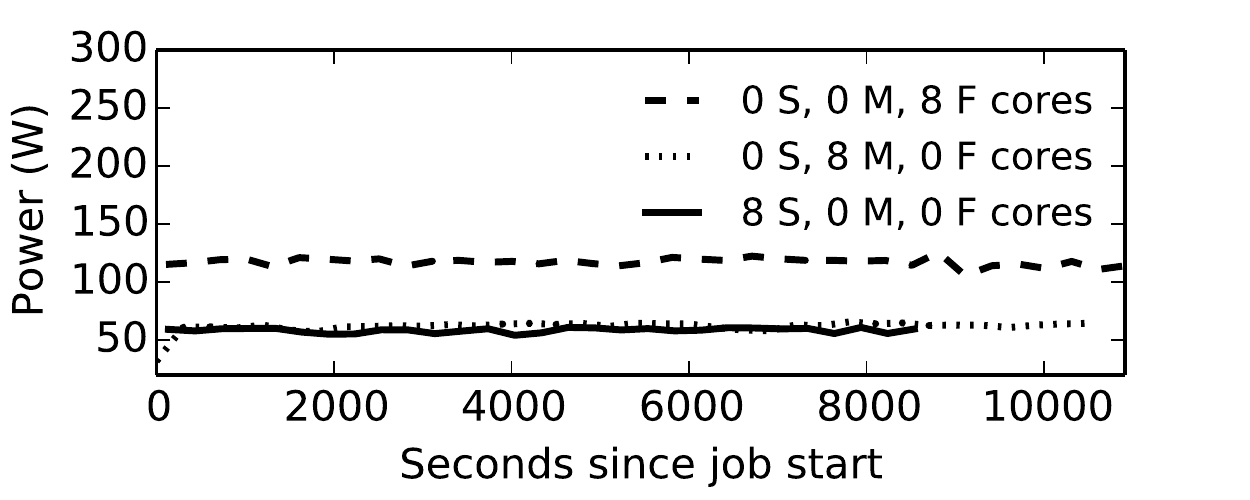}\caption{Three jobs with same name and same number of cores, but allocated on different classes of CPUs. }\label{fig_lowmedhigh}
    \end{subfigure}
     \hspace{0.1cm}
    \begin{subfigure}{0.48\textwidth}
     \vspace{-0.3cm}
    \includegraphics[width=\textwidth]{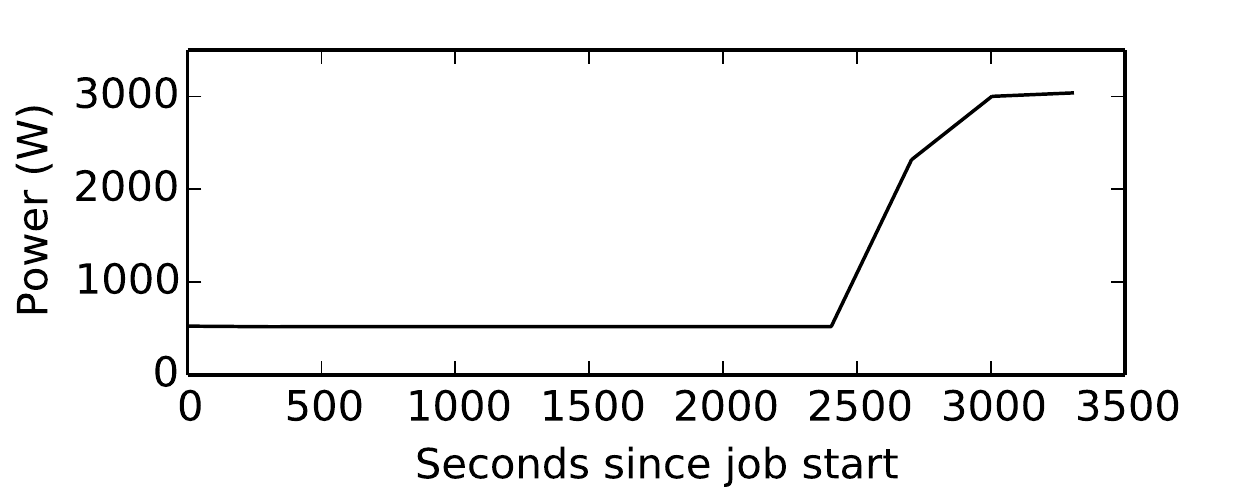}
    \caption{Power consumption of a single job throughout its execution.}\label{fig_runtime}
    \end{subfigure}
     \begin{subfigure}{0.48\textwidth}
    \includegraphics[width=\textwidth]{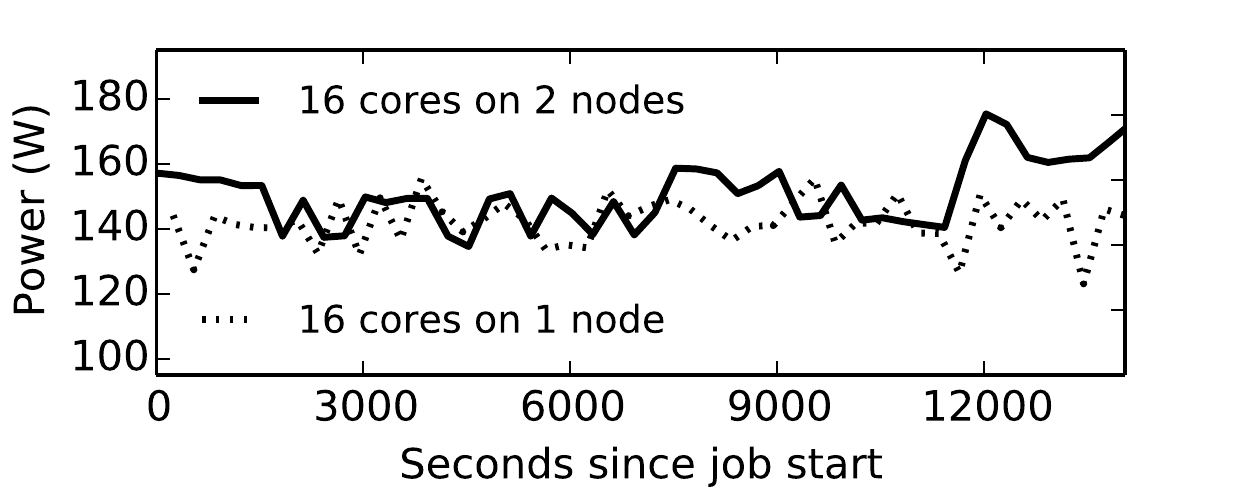}
    \caption{Two jobs with same name and same number of cores, but allocated on different number of nodes. }\label{fig_nnodes}
    \end{subfigure}
    \hspace{0.1cm}
     \begin{subfigure}{0.48\textwidth}
     \vspace{-0.15cm} 
    \includegraphics[width=\textwidth]{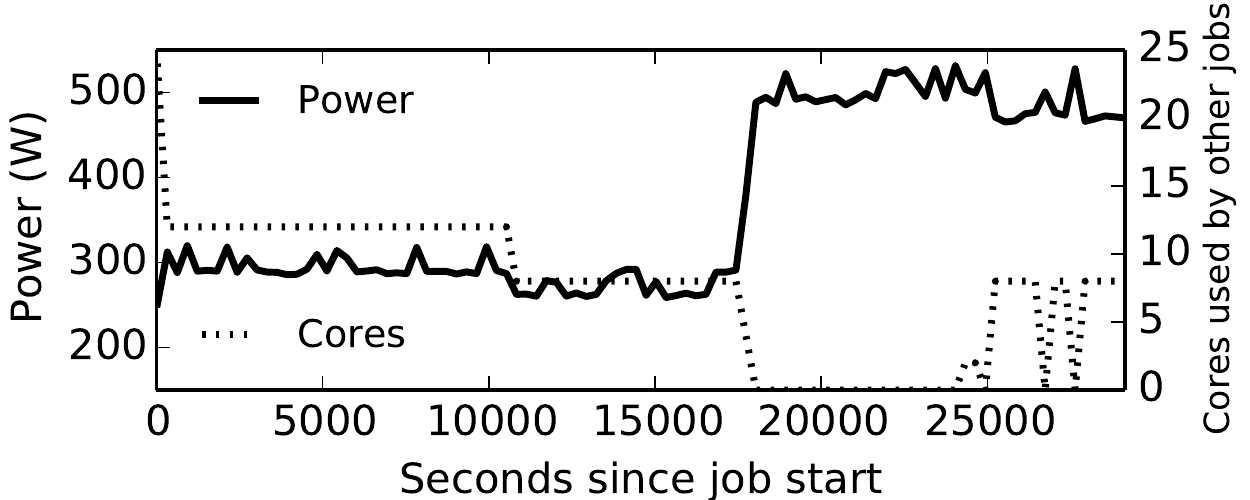}
    \caption{One job with variable number of cores in use by other users on the same nodes. The job itself uses 16 cores on 4 nodes. }\label{fig_shared}
    \end{subfigure}
    \caption{Power consumption for various jobs.}
  \end{center}
\end{figure}

\section{Power model}\label{model}
\subsection{Features}
Power consumption of a job can depend on several factors. 
One is the number of components of each type used by the job. A job using 16 cores will most likely use more power than a job using only 8 cores. A related factor is the type of core being used, with faster class cores using more power than slower class cores (e.g. Fig.~\ref{fig_lowmedhigh}). 
The structure of the application is also important. A job can have periods of high power usage and other periods of lower power usage, although the number of components in use remains the same (e.g. Fig.~\ref{fig_runtime}). These  patterns can be captured by including the runtime as a feature (i.e., the time since the job started). We also use the job name, a user-defined string, to identify the  application. We performed a textual analysis, using the \emph{CountVectorizer} class in the \emph{scikit-learn} python package \cite{scikit-learn},  checking which $n$-grams of 2 and 3 letters are present in the job names. This resulted in a set of numerical features that count how many times each $n$-gram appears.

The \emph{distribution} of resources is also important. Fig.~\ref{fig_nnodes} shows two jobs running the same application using 16 cores, however one job is allocated one node while the other two nodes, causing differences in power. A related factor is the load of a node that is partially used by a job. Fig. \ref{fig_shared} shows an example job using 16 cores on 4 nodes, together with the number of cores used by other jobs on the same nodes at the same time. We see a negative correlation between power and cores used by other jobs, so we include the same-node used cores as features.

To summarize, for each job we extracted the number of components of each type used ($S$, $M$, $F$, GPU, MIC), runtime, name (occurrence of $n$-grams of size 2 and 3), number of nodes and same-node components used by other jobs as \emph{regression features}, and the power as \emph{regression target}. These were computed at 5-minute intervals, resulting in numerous data points per job. Google BigQuery\cite{bigquery} was used, enabling analysis of large amounts of data in a reasonable amount of time.

\subsection{Regression problem and training procedure}
Since power is measured for each component type, we divided the problem of predicting power per job into 5 subproblems, corresponding to the terms on the right side of Eq.\ref{eq_pow}. Hence, for each user we perform 5 regression analyses, one for each component type, and then sum the predicted component powers to obtain an estimate for the \emph{global job power}. In practice, most users use only 2 or 3 component types, so regression is only performed for those. We use SVR with Radial Basis Function (RBF) kernels~\cite{svr}. 
To simulate the realistic scenario where power prediction is based on user histories, we train new models on a monthly basis using all past data, and then apply the models to new data for the current month. Here we show results for October 2014: models are trained on data prior to October 1st and then applied to all data recorded in October.

Training consists of two steps. First, SVR meta-parameters have to be optimized for each user and each component type. We use cross-validation to find optimal values. That is, all past data is divided into a ``train'' subset and a ``test'' subset. We use the first 80\% of the jobs of each user as training data, and the last 20\% as test data. Then multiple models are trained with a range of parameter values and the combination that produces the best results on the test data is selected.
Second, a new \emph{final model} is trained with the optimal parameter combination, using all past data in training (merging the train and test dataset).  This ensures that all available past information is included in the model. 

\begin{table}[b]
\centering
\begin{tabular}{|p{3cm}||p{1.3cm}|p{1.3cm}|p{1.3cm}|p{1.3cm}|p{1.3cm}|p{1.3cm}|}
\hline
Component type&$S$&$M$&$F$&GPU&MIC&Global\\
\hline
Users&21&20&27&9&2&34\\
\hline
\end{tabular}
\vspace{0.2cm}
\caption{Number of users analyzed for each component type and globally. }
\label{tab_users}
\end{table}

Once the final model is available for each user and component type, it can be applied for one month to new unseen data.  Power is predicted for the individual components and then summed to obtain \emph{global job power}.  To avoid poor prediction due to limited training, we only analyzed those users for which historical data included at least 1,000 total data points, at least 10 jobs totaling at least 100 time points for each component type, and at least 10 data points to apply the model to. Table \ref{tab_users} shows the number of users analyzed for each component type and globally. For all users, a total of 435,079 data points from 22,130 unique jobs were used to build the model (data before October), which was then applied to 53,717 new points from 5,039 unique jobs (October data).

\subsection{Evaluation}

We will compare results of our multiple-SVR model with a simple \emph{Enhanced Average Model} (EAM). For each user, the EAM computes the average power used per component unit (core, GPU, MIC), based on historic data. Then job power for each component is predicted by multiplying the number of component units by the average power per unit.  For instance, if a job $j$ of user $u$ uses $n^j_{F}$ $F$ cores and the historical usage for one $F$ core for the user is $\bar{P}^u_{F}$, then $P^{j}_{F}=n^j_{F} \times \bar{P}^u_{F}$. All other terms in Eq.~\ref{eq_pow} are computed in a similar fashion and then summed to obtain total job power. This model is an enhanced version of the so called ``average model'', which would compute job power just by averaging historical data, without taking into account the number of components used. Even so, it is much simpler than our multiple-SVR approach.  Training is straightforward as it only requires computing averages per user, while application of the model requires knowing only the number of components used.

The models were evaluated using two standard criteria for regression:  the (mean-)normalized-root-mean-squared-error (NRMSE) and  R-squared ($R^2$):
\begin{equation}
\makebox{NRMSE}=\frac{\sqrt{ (\sum_{i=1}^N{(P_i-P^*_i)^2}) / {N}}} {\bar{P}}
\end{equation}
\begin{equation}
R^2=1-\frac{\sum_{i=1}^N(P_i-P^*_i)^2}{\sum_{i=1}^N(P_i-\bar{P})^2}
\end{equation}
where $N$ is the number of data points considered across the jobs of the user, $P^*_i$ and $P_i$ are the predicted
and real powers for data point $i$, respectively,
while $\bar{P}$ is the average of the real power over all $N$  data points.

To provide context for the errors reported, it is important to understand the natural fluctuations of power consumption at constant load --- the noise levels. Power usage can vary for the same workload on the same node, due to \emph{hardware-related noise}, such as variations in the production process which may generate different electrical behaviors across same-type cores, or adaptive mechanisms for performance optimization~\cite{mccullough2011}.  Additionally, there is \emph{software-related noise}, introduced by operating system interference, external interrupts or shared resource contention~\cite{Fraternali2014}. Noise has a negative impact on power model performance since random fluctuations are not captured by model features, hence cannot be reproduced through regression. Thus, we cannot expect model errors to be less than the noise levels. This has been shown to \emph{affect performance of models} by reducing the maximum accuracy they can obtain~\cite{mccullough2011}.

\begin{figure}[!t]
\centering
\includegraphics[width=0.95\textwidth]{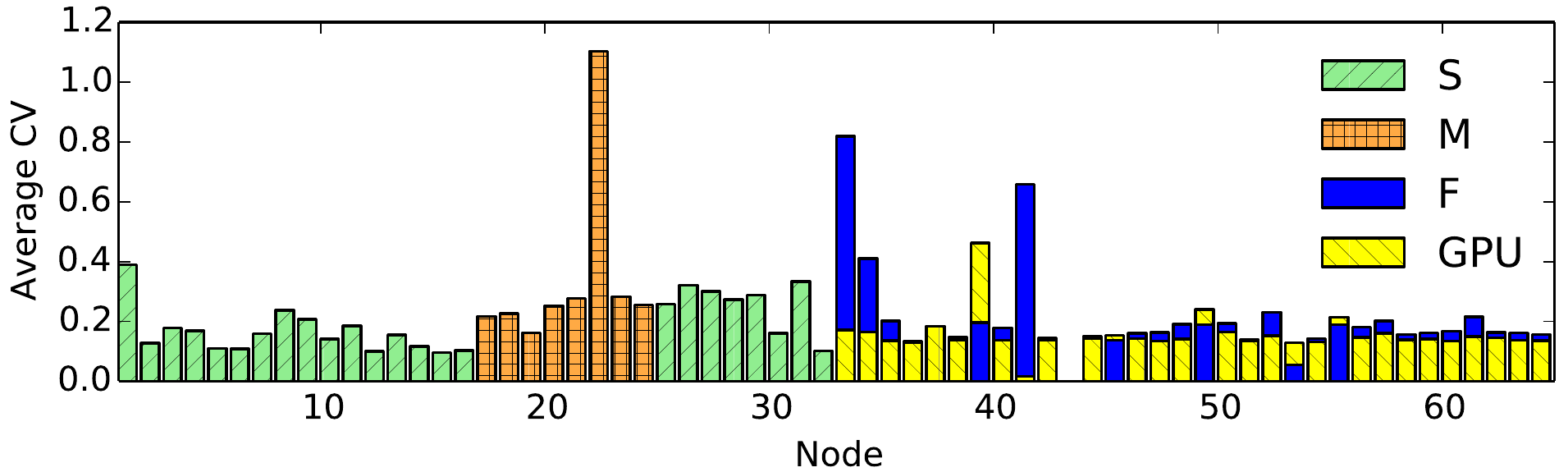}\\
\caption{Variability of power consumption. 
The bars represent the coefficient of variation (CV) of power at fixed load, averaged over all loads, for each node and component type in the system.}
\label{fig_var}
\end{figure}

In the case of {\Eurora}, undesired software and hardware variability \emph{between nodes} was previously shown to reach up to 20\% (5\% software and up to 15\% hardware)~\cite{Fraternali2014}. Here, we look at \emph{within-node variability} at constant load for CPUs and GPUs. Specifically, we computed the coefficient of variation (CV) of power at various load levels, and averaged over all loads for each component. For MICs, load information is not available so we could only analyze power at 0 load. 
\revised{The variability observed may come from different sources, however we evaluate them together since we are interested only in an overall value to be used as a baseline for quantifying our errors.}
Fig.~\ref{fig_var} shows average CV values for all 64 nodes, per component type. The $M$ CPUs show on average largest fluctuations, with most nodes reaching over 20\%. Most $S$, $F$ and GPU components have average fluctuations under 20\%. However, some nodes in all categories display much larger fluctuations, even over 100\%.  For MICs (not shown in the figure since load data is unavailable), idle power fluctuates on average by 10.24\% and we expect this value to be larger at larger loads. Hence, based on our data analysis and based on previous studies, in this work we consider NRMSE values  $<0.2$ (20\%)  to be good performance, since they are within the natural fluctuations of the individual components.

We included here both the $R^2$ and NRMSE evaluation criteria because they are complementary: the NRMSE looks at overall fit and gives a measure relative to the mean value, while $R^2$ looks at the general shape of the time series and gives a measure relative to the variations in the data. Additionally, they are affected differently by noise. 
For instance, if the power levels for a user are relatively flat, and vary only due to noise, the $R^2$ measure becomes irrelevant. This because $R^2$ looks at the `shape' of the data, which in this case is entirely determined by local fluctuations, which cannot be reproduced by any model. However, a model can still  capture average behavior which is the best performance possible, but which will correspond to low $R^2$. In this case the NRMSE  provides additional information, with a NRMSE value similar to the noise level considered a good performance. Conversely, when a user has highly variable power consumption for jobs, NRMSE can be large due to a few data points, but the model can still contain useful information, reflected in the $R^2$ measure.  In the following we will consider NRMSE $>0.2$ or  $R^2>0.5$ to be a very good result.

\begin{figure}[!t]
\centering
\includegraphics[width=0.9\textwidth]{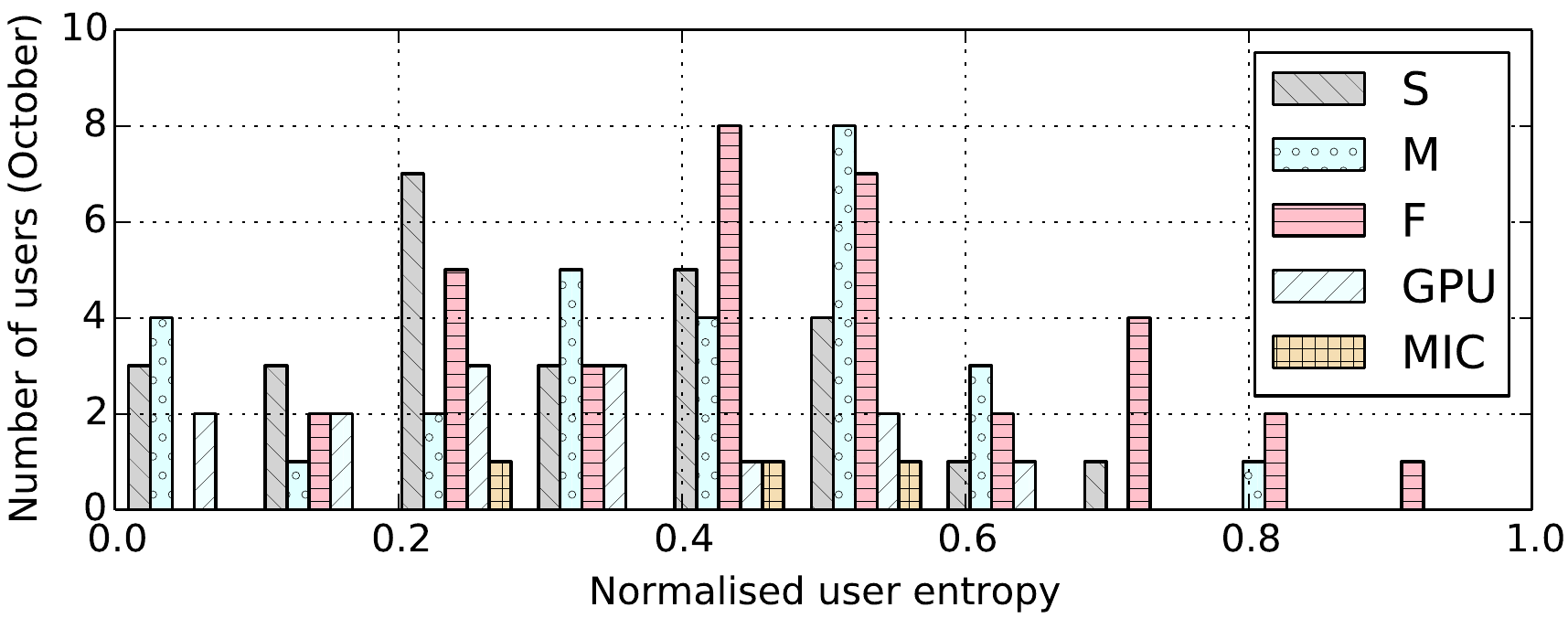}\\
\caption{\revised{Distribution of variability of power across jobs and time for each user. Variability for each user is computed as the normalized entropy of the distribution of job power levels recorded in time for that user. The plot shows a histogram of all entropies for the users active in the month of October 2014.} }
\label{fig_entropy}
\end{figure}

\revised{In our data, the distribution of job power for each user is very heterogeneous, with users ranging from those having jobs with stable power requirements to those showing very large differences across jobs and time, justifying the use of both NRMSE and $R^2$. Fig.~\ref{fig_entropy} shows the distribution of variability in job power for all users active during the month of October 2014, for each component type.  To quantify the variability for each user, we obtained the distribution of the power levels for that user, by collecting the data in bins of size 20W in the range [0W,500W], and computed the entropy of this distribution, normalized by the maximum entropy possible (logarithm of the number of bins). The normalized entropy is a measure of spread of the distribution, with 0 meaning all data fell into one bin and 1 meaning power levels are uniformly distributed across all bins. So, a user with 0 entropy has very flat power levels, while a user with high entropy has very large differences between power levels.  As Fig.~\ref{fig_entropy} shows, our data contains users with a wide range of entropies, hence we are dealing with a very heterogeneous user population.}
%

\section{Model performance}\label{global}

\begin{table}[b]
\small
\centering
\begin{tabular}{|p{2.2cm}||p{1.5cm}|p{1.5cm}|p{1.5cm}|p{1.5cm}|p{1.5cm}|}
\hline
&$S$&$M$&$F$&GPU&MIC\\
\hline
\hline
SVR NRMSE&0.13&0.33&0.52&0.15&0.28\\
\hline
SVR $R^2$&0.87&0.47&0.92&0.84&0.34\\
\hline
\hline
EAM NRMSE&0.13&0.37&1.34&0.24&0.28\\
\hline
EAM $R^2$&0.87&0.34&0.52&0.59&0.31\\
\hline
\end{tabular}
\vspace{0.2cm}
\caption{Performance of the SVR and EAM for individual components.}
\label{tab_results}
\end{table}

Once meta-parameters are explored using cross-validation with data prior to October 1st, 2014, the best meta-parameter combination is selected and a new final model is trained on all data prior to October. One SVR model is obtained for each user and each component type, which are then combined into a global model for each user. Table~\ref{tab_results} shows prediction performance for all users (all jobs concatenated), for each component type, throughout the month of October.
For components $S$ and GPU, both $R^2$ and NRMSE values are very good. For $F$, NRMSE is quite high, however $R^2$ is also very large, so the model contains useful information. The high NRMSE is due to the fact that one user has jobs that consume much more power than others (over 3KW versus under 500W for others), so a relatively small error in that user will produce a large overall NRMSE (due to the fact that the normalizing factor depends on all jobs of all users).
If we remove the user with jobs consuming over 3KW, then we obtain $R^2=0.89$ and $NRMSE=0.26$ which are very good considering the noise levels for the $F$ CPUs shown in Fig.~\ref{fig_var}. For $M$ CPUs, which showed highest noise in Fig.~\ref{fig_var}, performance is somewhat lower. $R^2$ does not reach the 0.5 threshold, albeit very close, while NRMSE is around 33\%. This shows how power fluctuations can affect model performance. Even so, the model is better than the average model ($R^2$ much larger than 0).
For the MIC component, the amount of data is more reduced, which can be one reason for the lower performance. Only two MIC users exist, one with very good  and another with lower prediction performance. 

\begin{figure}[!b]
\centering
\includegraphics[width=0.9\textwidth]{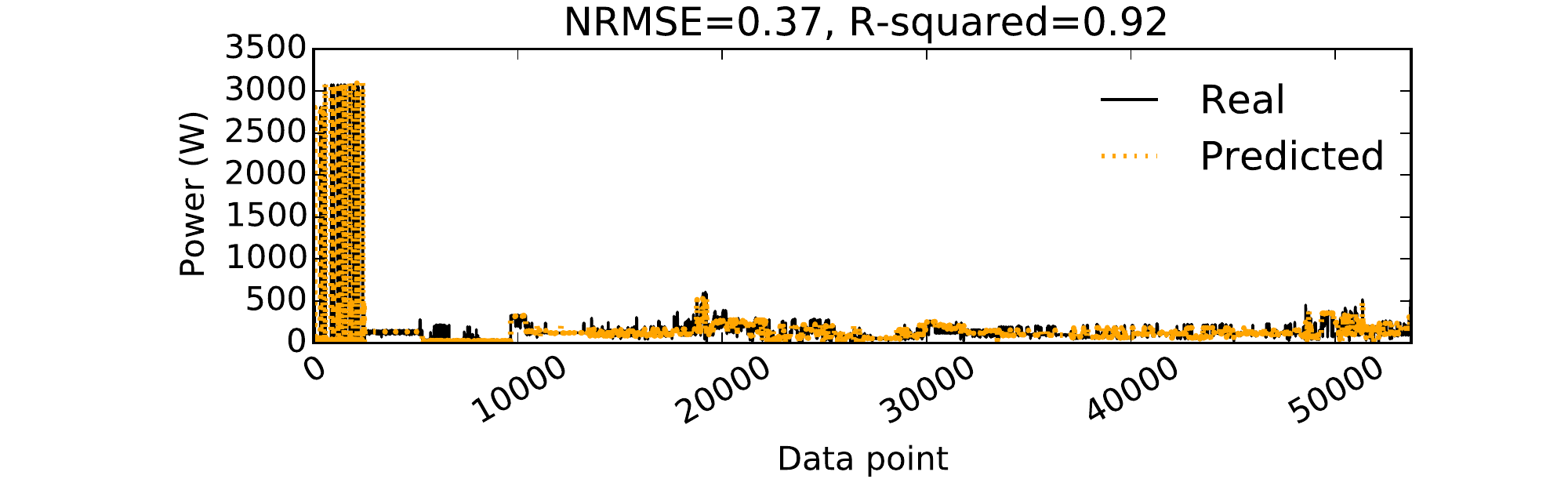}
\includegraphics[width=0.9\textwidth]{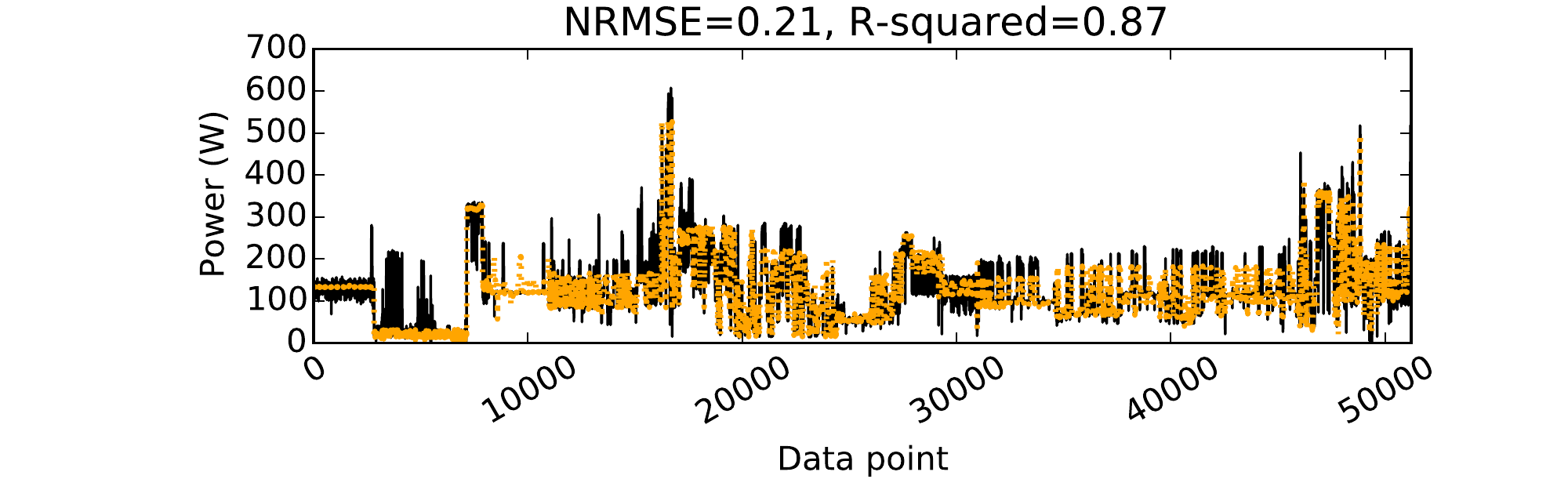}\\
\caption{Global real and predicted total power consumption (components summed together). For each job, the power
was computed at 5-minute intervals, with the plot showing all power values for all users and jobs. 
The top panel shows all users, while in the lower panel the first user with high power values was eliminated. }
\label{fig_total}
\end{figure}

If we compare the SVR models with the EAM, for which results are also shown in Table~\ref{tab_results}, we note that the SVR has better performance on all component types except for $S$. For the $S$ class, the SVR and EAM are comparable, meaning that jobs using this component are quite predictable and power depends mostly on the number of components used. Significantly better performance of the SVR can be seen for the $F$ and GPU components, which are also the most used across the cluster. This increase in performance means jobs are much more complex and additional SVR features are important in predicting the power outcome.

While Table~\ref{tab_results} shows how the model behaves on the individual components, it is total job power (global model) that interests us the most. Fig.~\ref{fig_total} shows the power time series (predicted and real) for the total job power (i.e., after applying Eq.~\ref{eq_pow}), using the SVR model.
Given the presence of that one user with very high job power, NRMSE is again large, however $R^2$ is very good. Again, by removing this user, NRMSE reduces to 21\%, meaning our model has an overall accuracy of 79\% for all other jobs of October, while $R^2$ stays high at 0.87. Compared to the EAM (global NRMSE$=0.91$ and R$^2=0.53$), NRMSE of
the SVR is 40\% that of the EAM, while $R^2$ is improved by 70\%.

\begin{figure}[!t]
\centering

\includegraphics[width=0.9\textwidth]{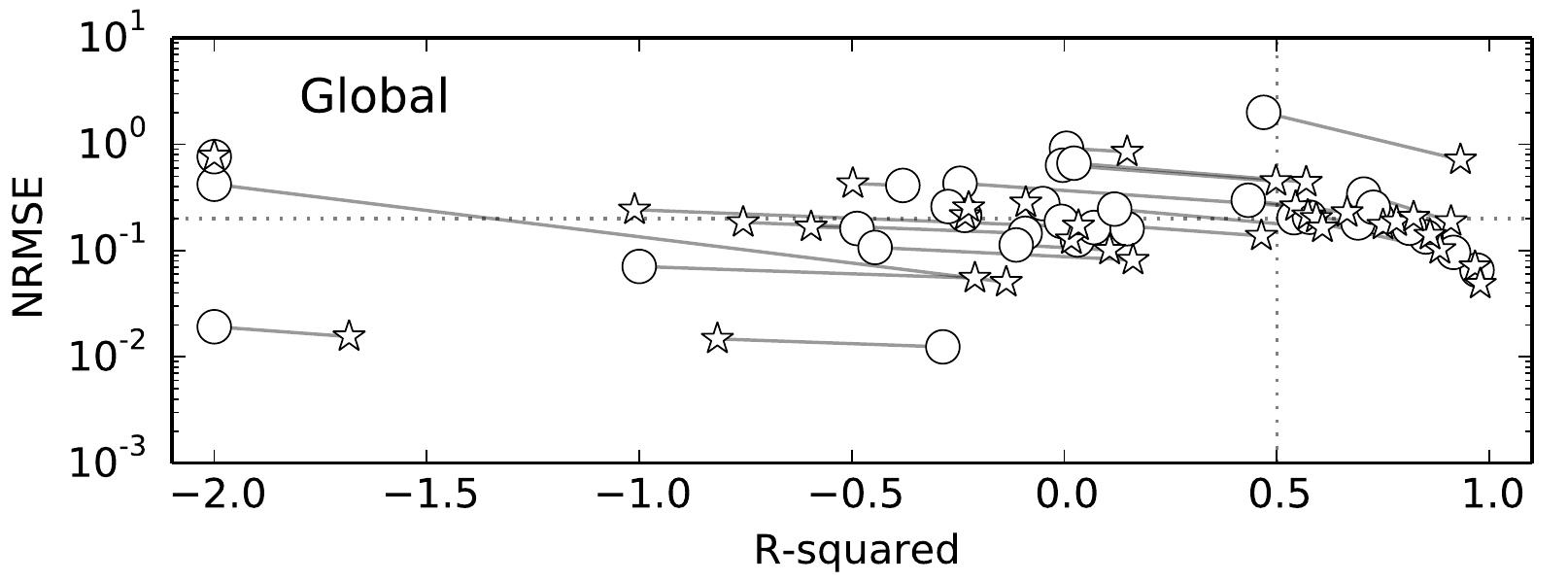}
\caption{Global model performance per user, NRMSE vs $R^2$. Circles show
performance for the EAM, while stars show performance for the multiple-SVR model. Each circle/star corresponds to one separate user. For each user, the two model types (EAM and SVR) are connected by an edge.
For figure readability, data points with very low negative $R^2$ values have been mapped to $-2.0$.}
\label{fig_users}
\end{figure}

Model performance varies also from user to user. Fig.~\ref{fig_users} plots global model NRMSE versus $R^2$ for each user, for both the SVR and EAM. In general, the SVR outperforms the EAM (in the plot, stars are located south-east of the corresponding circle), however there are a few users for which the EAM is better. For these, one is better off using the EAM for predictions. Out of a total of 34 users analyzed, 27 have SVR NRMSE $\leq0.2$ or $R^2\geq0.5$, and 7 (20\%) have lower performance. For the latter, a weak SVR model corresponds also to a weak  EAM model. Poor performance could be due to noise, indicated by the fact that jobs of these users use partial node resources (i.e., 1 out of 2 MICs or 1 out of 16 cores) or run on nodes with high variability, being thus more prone to noise.

\section{Related work}\label{related}

Power monitoring, modeling and optimization have been major research concerns in recent years. Modern computing units embed advanced control mechanisms such as the Dynamic Frequency and Voltage Scaling that aim to optimize performance and can affect  power levels, making modeling problematic even for a single computational unit~\cite{mccullough2011}. Several models trying to explain the relation between frequency, load, hardware counters and power for single units have been introduced for multicore CPUs~\cite{Dargie2015,Gschwandtner2014} and GPUs\cite{Ma2009}.  Model performance ranges widely depending on the applications running, with errors between 3.65 and 14.4\% for the CPU case, and between 1.7\% and 27.7\% for GPUs. These errors are only expected to grow when multiple units have to be combined, as is the case for HPC systems. Our approach is very different in that we are modeling power consumption per job, not per component. Additionally, our model does not require direct measures of load and frequency, which are typically not known in advance, but only workload measures which are known when the job starts.
 
Some work in modeling job or application power consumption has appeared recently. 
Performance counters are used to model application power on three small scale HPC platforms by~\cite{Witkowski2013}. GPU CUDA kernels are analyzed in~\cite{Nagasaka2010}, again based on job performance counters. These methods are very different from ours, since they require instrumenting the applications to extract signatures and performance counters, while we only use the number of resources required, making it much more straightforward to apply. An approach more similar to ours was recently introduced in~\cite{Shoukourian2014} which uses the number of nodes used by applications as model input, with very good precision (errors between 0 and 5.2\% per application). A more detailed model is introduced in~\cite{Storlie2014} where a Hidden Markov Model is used to represent job states and transitions.  All these methods build one model per job, while we are trying to explore user patterns as well. 
\revised{Building one model per user has several advantages including larger training datasets and greater robustness to inaccurate use of job names by HPC users (e.g., when the user gives the same job name to different applications, or various names to the same application).}
Additionally, unlike other methods, our model applies to hybrid jobs using CPUs, GPUs and MICs.

On the road towards ExaFLOP performance, special attention has been given to system-level power consumption by clusters. Recent work at Google~\cite{Gao2014} describes the use of Artificial Neural Networks to model Power Usage Effectiveness using a mixture of workload and cooling features. 
System-level prediction of power consumption is also one application of our predictive model .
In terms of power-aware scheduling, another possible application of our models, the authors in~\cite{Borghesi2015,Borghesi2015a} introduce a method based on Constraint Programming, to achieve power capping on {\Eurora}, the same HPC system analyzed here. This could benefit greatly from power prediction offered by our framework.


\section{Discussion and Conclusions}\label{discussion}

We presented an analysis of \emph{historical trace data} from {\Eurora}, and evaluated prediction models for power consumption of jobs. The method is fully data driven --- no assumptions about the model structure nor additional instrumentation of application code are required. 
\revised{The only application-aware feature is the job name, making our method easily applicable to any system even when application code is not available. The power of our prediction derives from user history rather than from application counters, and our results show that when enough data is available, high performance can be achieved.}
We employ a multiple-SVR model to estimate job power in time. One model per user is trained.
\revised{An alternative would have been to build one model per application (job name) but this would have meant much less training data per model.}
\revised{Additionally, learning from user profiles can allow for user trends to be captured, maintaining high quality predictions even if job names are not properly employed by users (e.g., using the same name for different applications or many different names for the same application).}

The multiple-SVR approach is compared to an enhanced average model (EAM) where power depends only on the number of components used. The SVR outperforms the EAM approach for most users, obtaining good prediction (error under 20\% or $R^2\leq0.5$) for 80\% of the users analyzed.  For the rest of the users, indications are that performance is affected by noise.

The approach is intended to be used in \emph{real time}, where predictions are made as new jobs arrive at the scheduler. Online application consists of training the model for each user, then applying it to real time data, by employing the procedure outlined in this work. Periodically, the model is updated by incorporating recent data into the training dataset. We expect monthly model updates to be sufficient in order to capture changes in job structure.  
available, prediction can be improved by training the multiple-SVR model. 

In terms of resources, our analysis was performed on a 516-node CentOS 7.0 cluster, with 2 octa-core 2.40GHz Intel Xeon CPUs per node. Since our problem is intrinsically parallel, we obtained each model separately on one core. Running times depended on the user (different amounts of data available) and on the meta-parameters. 
\emph{Meta-parameter optimization} required a total of 185.36 core-hours for all users, with a maximum running time for one optimization run of 4.66 hours. \emph{Global model training} for all users required at total of 2.92 core-hours (maximum for one user was 1.7 hours), while \emph{model application} to all the data of October took only 6.81 minutes for all users.  Consequently, if parallelized on a multi-core platform, the entire process incurs little overhead, especially given that the training procedure has to be repeated only once a month.
\revised{We expect the method to scale to systems that are much larger than Eurora, since the analysis is performed separately for each user and can be easily parallelized.}

The predictions presented here can be  \revised{improved through more detailed data on job characteristics (e.g., exact application names, input datasets and parameters) and more detailed power monitoring (e.g., power per core rather than per CPU), work which we will undertake in the future after obtaining improved datasets. Furthermore, we plan to use our predictions }
 in several applications to optimize system functionality. The first is modeling and prediction of system level power consumption, including networking equipment, IO systems and even cooling infrastructure, starting from prediction of job power. Secondly, our approach is applicable to power-aware scheduling, where the scheduler can estimate power usage for various job allocation schemes and select the best among them. Thirdly, our method can be employed by users to estimate power for their jobs before submission, which can facilitate better management of resources by the users, especially in the context of power-aware billing.


\section{Acknowledgments}
BigQuery analysis was carried out through a generous Cloud Credits grant from Google. We are grateful to Prof. L. Benini and Dr. A. Bartolini for useful discussions regarding the data and to the HPC group at  {\Cineca}, in particular Dr. E. Rossi and Dr. C. Cavazzoni for providing access to the {\Cineca} systems. We acknowledge the {\Cineca} ISCRA  PACNA and PM-HPC  awards allowing access to HPC resources and support. \revised{This work was partially funded by the European project SoBigData Research Infrastructure --- Big Data and Social Mining Ecosystem under the INFRAIA-H2020 program (grant agreement 654024).}

\bibliographystyle{splncs03}
\bibliography{refs} 

\end{document}